\documentclass[oldversion]{aa}
\usepackage{epsfig}
\usepackage{natbib}
\usepackage{lscape}
\usepackage{wasysym}
\bibpunct{(}{)}{;}{a}{}{,}
\begin{document}

\title{ELODIE metallicity-biased search for transiting Hot Jupiters\thanks{Based on 
observations collected at the La Silla-Paranal Observatory,
ESO (Chile) with the CORALIE spectrograph at the Euler 1.2-m Swiss Telescope, and
with the ELODIE and SOPHIE spectrographs at the 1.93-m telescope of the Observatoire de Haute 
Provence (OHP), France.}}

\subtitle{V. An intermediate-period Jovian planet orbiting HD\,45652}

\author{
  N. C. Santos\inst{1,2} \and		
  S. Udry\inst{2} \and 
  F. Bouchy\inst{3} \and 
  R. Da Silva\inst{2} \and 
  B. Loeillet\inst{3,4} \and 
  M. Mayor\inst{2} \and 
  C. Moutou\inst{4} \and 
  F. Pont\inst{2} \and 
  D. Queloz\inst{2} \and 
  S. Zucker\inst{5} 
  D. Naef\inst{2,6} \and 
  F. Pepe\inst{2} \and 
  D. S\'egransan\inst{2} \and 
  I. Boisse\inst{3} \and 
  X. Bonfils\inst{1,7,8} \and 
  X. Delfosse\inst{7} \and 
  M. Desort\inst{7} \and 
  T. Forveille\inst{7} \and 
  G. H\'ebrard\inst{3} \and 
  A.-M. Lagrange\inst{7} \and 
  C. Lovis\inst{2} \and 
  C. Perrier\inst{7} \and 
  A. Vidal-Madjar\inst{3} 
  }

\institute{
    Centro de Astrof{\'\i}sica, Universidade do Porto, Rua das Estrelas, 
    4150-762 Porto, Portugal
    \and
    Observatoire de Gen\`eve, 51 ch. des Maillettes, 1290 Sauverny, Switzerland
    \and
    Institut d'Astrophysique de Paris, UMR7095 CNRS, Universit\'e Pierre \& Marie Curie, 
    98bis Bd Arago, 75014 Paris, France 
    \and
    Laboratoire d'Astrophysique de Marseille, Traverse du Siphon, 13376 Marseille 
    Cedex 12, France
    \and
    Department of Geophysics and Planetary Sciences, Raymond and Beverly Sackler Faculty 
    of Exact Sciences, Tel Aviv University, Tel Aviv 69978, Israel
    \and
    European Southern Observatory, Casilla 19001, Santiago 19, Chile
    \and
    Laboratoire d'Astrophysique, Observatoire de Grenoble, UJF, CNRS, BP 53, 38041 Grenoble Cedex 9, France
    \and
    Centro de Astrof{\'\i}sica da Universidade de Lisboa, Observat\'orio Astron\'omico de Lisboa,
    Tapada da Ajuda, 1349-018 Lisboa, Portugal
}


\date{Received XXX; accepted XXX}

\abstract{
We present the detection of a 0.47\,M$_{\mathrm{Jup}}$ planet 
in a 44-day period eccentric trajectory ($e$=0.39) orbiting the metal-rich
star \object{HD\,45652}. This planet, the seventh giant planet discovered in the context of
the ELODIE metallicity-biased planet search program, is also confirmed using 
higher precision radial-velocities obtained with the CORALIE and SOPHIE spectrographs. 
The orbital period of HD\,45652b places it in the middle of the ``gap'' in 
the period distribution of extra-solar planets.
  \keywords{stars: individual: HD\,45652 -- 
            stars: planetary systems --
	    planetary systems: formation -- 
            techniques: radial-velocity -- 
	    stars: fundamental parameters
	    }}

\authorrunning{Santos et al.}
\titlerunning{ELODIE metallicity-biased search for transiting Hot Jupiters V.}
\maketitle

\section{Introduction}

More than 280 exoplanets have been found orbiting solar-type stars, most of them
discovered using radial-velocity techniques 
\citep[for a review see][]{Udry-2007}\footnote{See updated tables at http://www.exoplanets.eu and http://www.exoplanet.eu}.
In about 40 cases, the discovered planets are known to transit their
host stars. The detection of a photometric signature of
a transiting planet, when complemented with its dynamical detection 
by means of the radial-velocity technique, provides the possibility of deriving its
mass, radius, and mean density \citep[e.g.][]{Charbonneau-2000,Pont-2005b}. These data provide invaluable information about the physical properties 
of the planet \citep[e.g. its composition --][]{Valencia-2006,Fortney-2007}, 
as well as about its formation and evolution.

It is well known that the probability of finding a giant planet is a strongly
rising function of stellar metallicity \citep[][]{Gonzalez-1998,Santos-2004b,Fischer-2005}. 
This correlation is generally accepted to reflect the higher probability of forming planets
around stars with a higher dust content in the proto-planetary disk, and supports
the core-accretion model for giant planet 
formation \citep[e.g.][]{Ida-2004b,Benz-2006,Matsuo-2007}\footnote{See however recent 
discussion in \citet[][]{Pasquini-2007} and \citet[][]{Hekker-2007}}.

Using the metallicity-giant planet relation, several programs
are now searching for planets around high metal-content stars 
\citep[e.g.][]{Tinney-2002,Fischer-2005b,DaSilva-2006,Melo-2007}. These programs 
mostly unveiled 
short-period planets (as expected due to their observing strategy), 
and increased the number of detected transiting planets orbiting bright 
stars \citep[e.g.][]{Sato-2005,Bouchy-2005b}.

One of these programs was based on the former ELODIE spectrograph \citep[][]{Baranne-1996}, 
mounted on the 1.93-m telescope at the OHP observatory \citep[for details of the program see][]{DaSilva-2006}. 
The results of this survey were presented in a series of four papers
\citep[][]{DaSilva-2006,Bouchy-2005b,Moutou-2006,DaSilva-2007}, announcing the discovery
of 6 giant planets, one of which transits its star (\object{HD\,189733b}). 

In this paper, we announce the discovery of a $\sim$0.5 Jupiter-mass companion orbiting 
\object{HD\,45652}, one star from the ELODIE metallicity-biased survey. 
In Sect.\ref{sec:star}, we describe the stellar
characteristics of HD\,45652. In Sect.\ref{sec:planet}, we present the radial-velocity
measurements used to detect HD\,45652b, as well as the derived orbital solution and
planetary characteristics. We present our conclusions in Sect.\,\ref{sec:conclusion}, discussing
how this new intermediate-period planet is placed in the general picture of giant-planet formation models.

\section{Stellar properties of HD\,45652}
\label{sec:star}

According to the SIMBAD database, \object{HD\,45652} (HIP\,30905, BD\,$+$11\,1197) is a high 
proper-motion V=8.1, K5 star (B$-$V=0.85). In the Hipparcos catalogue \citep[][]{ESA-1997},
the star is listed as having a parallax of $\pi$=27.67$\pm$1.29\,mas, 
a value that implies a distance of 36$\pm$2\,pc. No reference for a close companion 
to \object{HD\,45652} is mentioned in this catalogue, and no significant photometric variability was detected. 

\begin{figure}[t]
\resizebox{\hsize}{!}{\includegraphics{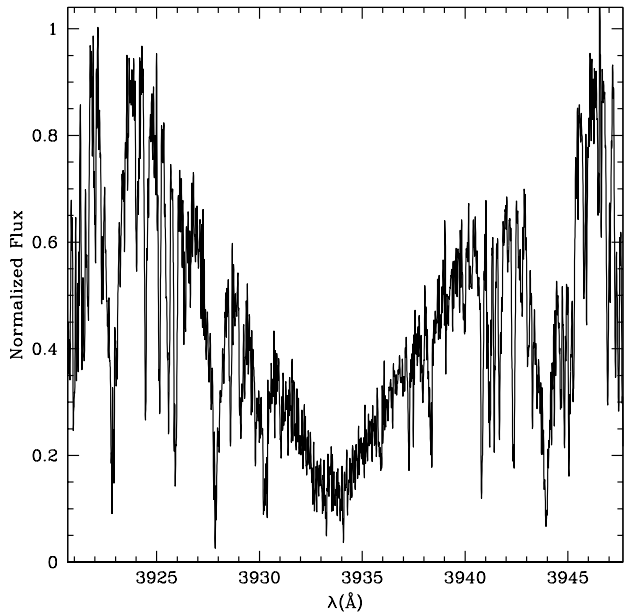}}
\caption{Spectral region of HD\,45652 centered at the \ion{Ca}{ii} K line (CORALIE spectrum). } 
\label{fig:caii}
\end{figure}

We used a high-resolution (R=50\,000) CORALIE spectrum with a signal-to-noise-ratio (S/N) $\sim$100 to derive
the stellar parameters and metallicity of HD\,45652, using the methodology described
in \citet[][]{Santos-2004b}. The resulting effective temperature, surface gravity, and [Fe/H] are 5312\,K, 4.32\,dex,
and $+$0.29\,dex, respectively (see Table\,\ref{table:hd45652_star}). These parameters 
are typical of a metal-rich main-sequence late-G or early-K dwarf, and slightly disagree with
the spectral type of K5 listed in the Hipparcos catalogue \citep[][]{ESA-1997}
and in SIMBAD. The above-mentioned temperature is also in good agreement with the value expected on the basis of its B$-$V colour 
and metallicity \citep[5315\,K, using the calibration presented in][]{Santos-2004b}, 
as well as with other temperature estimates in the literature, namely
by \citet[][T$_\mathrm{eff}$=5150\,K]{Strassmeier-2000}, \citet[][T$_\mathrm{eff}$=5370\,K]{Allende-1999}, and \citet[][T$_\mathrm{eff}$=5349\,K]{Robinson-2007}.

Using stellar evolution tracks from \citet[][]{Schaerer-1993a}, we derived a
stellar mass of 0.83$\pm$0.05\,M$_\odot$, and an age above 12\,Gyr for HD\,45652.
\citet[][]{Strassmeier-2000} classified this star as chromospherically non-active,
in agreement with its derived age. From SOPHIE spectra, we also derived a value of
$\log{R'_{HK}}=-4.90\pm0.10$. The low activity level is confirmed by an inspection of the 
center of the \ion{Ca}{ii} H and K line
regions (see Fig.\,\ref{fig:caii}). Such a low level of activity is also compatible
with the low value of projected rotational velocity ($v\,\sin{i}<2$\,km\,s$^{-1}$) as derived from the
Cross-Correlation Function (CCF) of the ELODIE spectra (Table\,\ref{table:hd45652_star}).

\begin{table}[b]
\caption{Stellar parameters for \object{HD\,45652}.}
\label{table:hd45652_star}
\begin{tabular}{lcc}
\hline\hline
\noalign{\smallskip}
Parameter  & Value & Reference \\
\hline
Spectral~type                   & K5/G8-K0              & Hipparcos/This Paper \\
$m_v$                           & 8.1                   & Hipparcos \\
$B-V$                           & 0.85                  & Hipparcos \\
Distance~[pc]                   & 36$\pm$2              & Hipparcos \\
$v~\sin{i}$~[km~s$^{-1}$]       & $<$2$\dagger$  	& This paper \\
$\log{R'_\mathrm{HK}}$	        & $-$4.90$\pm$0.10$\dagger\dagger$      & This paper \\
$T_{\rm eff}$~[K]               & 5312$~\pm~$68         & This paper \\
$\log{g}$~[cgi]                 & 4.32$~\pm~$0.21       & This paper \\
$\xi_{\mathrm{t}}$              & 0.89$~\pm~$0.09       & This paper \\
${\rm [Fe/H]}$                  & $+$0.29$~\pm~$0.07    & This paper \\
Mass~$[M_{\odot}]$              & 0.83$\pm$0.05         & This paper \\
\hline
\noalign{\smallskip}
\end{tabular}
\newline
$\dagger$ From ELODIE spectra using a calibration similar to that presented by \citet{Santos-2002a}\\
$\dagger\dagger$ From SOPHIE spectra using a calibration similar to that presented by \citet{Santos-2000a}
\end{table}

\begin{figure}[t]
\resizebox{\hsize}{!}{\includegraphics{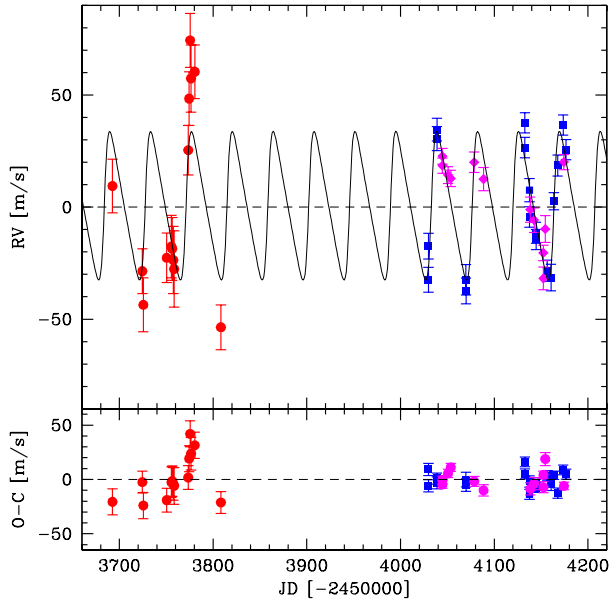}}
\caption{{\it Top}: ELODIE (red circles), CORALIE (blue squares), and SOPHIE (magenta diamonds) 
radial-velocities for HD\,45652 as a function of time, and the best-fit Keplerian function to the 
data. {\it Bottom}: residuals of the fit.} 
\label{fig:time}
\end{figure}

\section{A giant planet orbiting HD\,45652}
\label{sec:planet}

\object{HD\,45652} was part of the ELODIE metallicity-biased planet search 
sample \citep[][]{DaSilva-2006}. A series of 14 radial velocities were obtained
with this instrument between October 2005 and March 2006 (Table\,\ref{table:rv}),
using the 1.93-m telescope at the Observatoire de Haute Provence (France). 
The data showed the presence of a clear radial-velocity variation, although the time
coverage of the data did not allow us to confirm the nature of the observed
signal. The average photon-noise error of the measurements is 12.6\,m\,s$^{-1}$.

\begin{figure}[t]
\resizebox{\hsize}{!}{\includegraphics{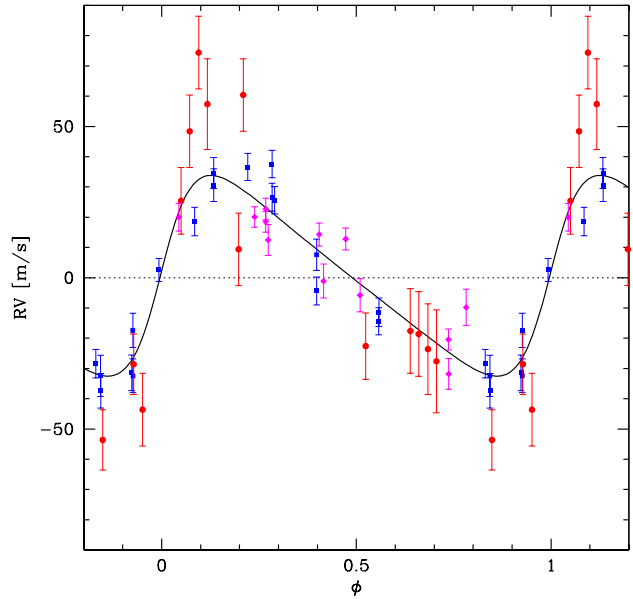}}
\caption{Phase-folded ELODIE, CORALIE, and SOPHIE radial velocities for HD\,45652 at
a period of 43.6\,days. Symbols are as in Fig.\,\ref{fig:time}.} 
\label{fig:phase}
\end{figure}

HD\,45652 was then monitored using the CORALIE spectrograph, at the 
Euler 1.2-m Swiss telescope (La Silla observatory, ESO, Chile -- between October 2006 and March 2007), and with the SOPHIE spectrograph, at the
1.93-m OHP telescope (France -- between November 2006 and March 2007).
The average photon-noise error of the 18 CORALIE and 12 SOPHIE radial-velocities is 5.5\,m\,s$^{-1}$ and 4.4\,m\,s$^{-1}$, respectively. 
Both sets of higher quality radial-velocities obtained (Table\,\ref{table:rv}) confirm
our earlier results, showing a clear radial-velocity signal (Fig.\,\ref{fig:time}).

The combined ELODIE, CORALIE, and SOPHIE radial-velocity measurements are best fited using 
a Keplerian function with a semi-amplitude $K$=33\,m\,s$^{-1}$, an 
eccentricity $e$=0.38, and a period $P$=43.6\,days (see Figs.\,\ref{fig:time} and 
\ref{fig:phase}, and Table\,\ref{table:hd45652_orbit}). The residuals
around the fit (8.9\,m\,s$^{-1}$) are within the expected value
given the photon-noise errors of the data. Monte Carlo simulation results
shows that the false-alarm probability is only 0.1\%. The observed 
Keplerian fit corresponds 
to the expected radial-velocity variation induced by the presence of a giant 
planet with a minimum mass of 0.47\,M$_\mathrm{Jup}$, orbiting HD\,45652 in an
eccentric orbit with a semi-major axis of 0.23\,AU.

\begin{table}[b]
\caption[]{Elements of the fitted orbit for HD\,45652b, derived using the 
ELODIE, CORALIE, and SOPHIE data.}
\begin{tabular}{lll}
\hline
\hline
\noalign{\smallskip}
$P$             & 43.6$\pm$0.2                          & [d]\\
$T$             & 2454120.3$\pm$1.2                     & [d]\\
$a$             & 0.23                                  & [AU]\\
$e$             & 0.38$\pm$0.06                         &  \\
$V_r$ (ELODIE)  & $-$5.117$\pm$0.006                    & [km\,s$^{-1}$]\\
$V_r$ (CORALIE) & $-$5.044$\pm$0.002                    & [km\,s$^{-1}$]\\
$V_r$ (SOPHIE)  & $-$4.999$\pm$0.002                    & [km\,s$^{-1}$]\\
$\omega$        & 273$\pm$12                            & [degr] \\ 
$K_1$           & 33.1$\pm$2.5                          & [m\,s$^{-1}$] \\
$\sigma(O-C)$   & 8.9                        	        & [m\,s$^{-1}$]  \\    
$N$             & 44                                    &  \\
$m_2\,\sin{i}$  & 0.47                                  & [M$_{\mathrm{Jup}}$]\\
\noalign{\smallskip}
\hline
\end{tabular}
\newline
\label{table:hd45652_orbit}
\end{table}

A similar orbital solution is obtained if we use only the CORALIE and SOPHIE data to derive 
the best-fit Keplerian function (similar orbital parameters are obtained, within the errors). 
The observed residuals are smaller in this case (7.4\,m\,s$^{-1}$).

\begin{table}[h]
\caption{ELODIE and CORALIE radial-velocity measurements of HD\,45652. }
\label{table:rv}
\begin{tabular}{ccc}
\hline\hline
JD     & $V_r$ [km\,s$^{-1}$] & $\sigma(V_r)$ [km\,s$^{-1}$]\\
\hline
\multicolumn{3}{l}{ELODIE}\\
2453692.65990 & $-$5.108 & 0.012 \\
2453724.51830 & $-$5.146 & 0.010 \\
2453725.52840 & $-$5.161 & 0.012 \\
2453750.52140 & $-$5.140 & 0.011 \\
2453755.52560 & $-$5.135 & 0.014 \\
2453756.46230 & $-$5.136 & 0.014 \\
2453757.49900 & $-$5.141 & 0.015 \\
2453758.44960 & $-$5.145 & 0.017 \\
2453773.45970 & $-$5.092 & 0.011 \\
2453774.42840 & $-$5.069 & 0.012 \\
2453775.42810 & $-$5.043 & 0.012 \\
2453776.41300 & $-$5.060 & 0.015 \\
2453780.42370 & $-$5.057 & 0.012 \\
2453808.29880 & $-$5.171 & 0.010 \\
\multicolumn{3}{l}{CORALIE}\\
2454029.79264 & $-$5.0607 & 0.0063 \\
2454029.80422 & $-$5.0757 & 0.0063 \\
2454038.83579 & $-$5.0093 & 0.0057 \\
2454038.84718 & $-$5.0125 & 0.0060 \\
2454069.78922 & $-$5.0759 & 0.0076 \\
2454069.81454 & $-$5.0807 & 0.0063 \\
2454132.62897 & $-$5.0005 & 0.0050 \\
2454132.64037 & $-$5.0107 & 0.0050 \\
2454137.62078 & $-$5.0233 & 0.0056 \\
2454137.63218 & $-$5.0301 & 0.0050 \\
2454144.58745 & $-$5.0442 & 0.0049 \\
2454144.59885 & $-$5.0368 & 0.0052 \\
2454151.59608 & $-$5.0445 & 0.0048 \\
2454156.52101 & $-$5.0726 & 0.0052 \\
2454160.52499 & $-$5.0747 & 0.0066 \\
2454163.57158 & $-$5.0404 & 0.0044 \\
2454167.58467 & $-$5.0250 & 0.0053 \\
2454173.53990 & $-$5.0061 & 0.0050 \\
2454176.55337 & $-$5.0187 & 0.0052 \\
\multicolumn{3}{l}{SOPHIE}\\
2454044.67360 & $-$4.9763 & 0.0035 \\
2454044.68380 & $-$4.9805 & 0.0034 \\
2454050.66630 & $-$4.9847 & 0.0038 \\
2454053.63090 & $-$4.9862 & 0.0036 \\
2454078.56860 & $-$4.9790 & 0.0045 \\
2454088.56780 & $-$4.9865 & 0.0051 \\
2454138.40400 & $-$5.0001 & 0.0056 \\
2454142.48580 & $-$5.0048 & 0.0055 \\
2454152.40650 & $-$5.0194 & 0.0034 \\
2454152.41540 & $-$5.0308 & 0.0051 \\
2454154.37280 & $-$5.0088 & 0.0059 \\
2454174.29350 & $-$4.9789 & 0.0033 \\
\hline				 
\end{tabular}			 
\end{table}			     

A look at the residuals of the Keplerian function fitting (Fig.\,\ref{fig:time}) suggests that
the adopted orbital solution does not well describe the ELODIE data.
The observed residuals are probably due to the higher quality 
CORALIE and SOPHIE measurements which are then weighted more signiticantly than the 
ELODIE measurements when fitting the data. 
Since we do not have any overlap 
between the ELODIE and CORALIE datasets, it may be that the
zero point of radial-velocity measurements for the different instruments is not well constrained.
Alternatively, a second signal could be present in the data. A scrutiny of old CORAVEL 
data, which has a precision of $\sim$300\,m\,s$^{-1}$,
does not uncover the existence of any long-term radial-velocity trend. 
We also studied the residuals of the
Keplerian fit for the presence of a second significant low-amplitude signal in the data, 
without success. We thus do not endorse the idea that a second signal
is present.

\begin{figure}[t]
\resizebox{\hsize}{!}{\includegraphics{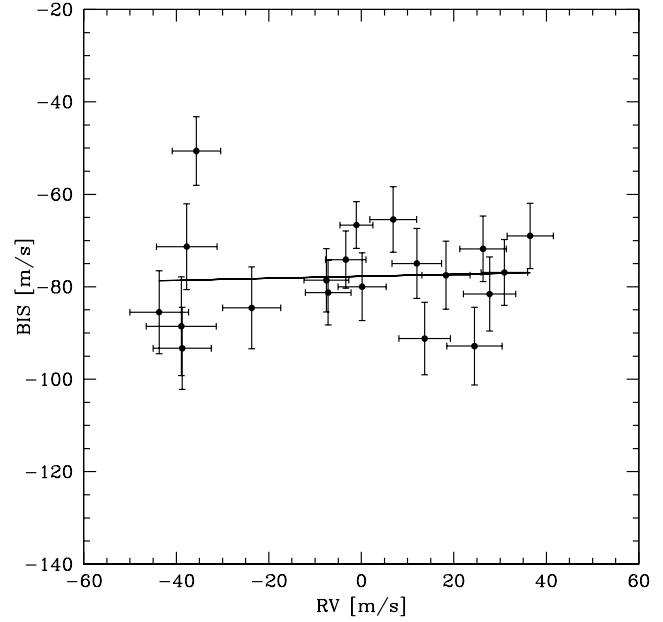}}
\caption{BIS versus radial-velocity for the CORALIE data of HD\,45652. To 
demonstrate the nonexistence of a correlation, the vertical and 
horizontal scales were set to be identical. The best linear fit to the data
is also shown. The slope of the fit has a non significant value of
0.02$\pm$0.09.} 
\label{fig:bisvr}
\end{figure}

Given the low projected rotational velocity and activity level of the star, 
we do not expect that activity-related phenomena could induce the observed periodic
radial-velocity signal. Although not a fully effective diagnostic for stars rotating very slowly \citep[][]{Santos-2003b,Desort-2007}, an analysis of the CCF bisectors
also lends support to this idea (Fig.\,\ref{fig:bisvr}). No significant 
correlation is found between the observed radial-velocity and the shape of the CCF 
denoted by a measurement of the Bisector Inverse Slope, BIS \citep[defined as in][]{Queloz-2001}.
The observed radial-velocity variation of HD\,45652 is therefore best explained 
by the presence of a planetary-mass companion in a $\sim$44-day period orbit.

\section{Discussion and concluding remarks}
\label{sec:conclusion}

Statistical studies of the properties of observed extra-solar planets
have shown that the orbital period distribution is characterized by
a well-defined peak for $P<$10-days, followed by some sort of
period valley for values of $P$ up to $\sim$100\,days. Above this latter value, 
the distribution is again an increasing function of orbital 
period \citep[][]{Cumming-1999,Udry-2003,Udry-2007}. The observed shape of
the period distribution is predicted by some models of planet formation
and evolution \citep[][]{Ida-2004a,Benz-2007}. 

With an orbital period of $\sim$44-days, HD\,45652b is placed in the middle of
the so-called period valley. 

Interestingly, \citet[][]{Burkert-2007} pointed out that the
observed period valley is more pronounced for planets orbiting more massive
stars (F-dwarfs with mass above $\sim$1.2\,M$_\odot$). Furthermore, the period
gap appears to be more significant for the higher mass planets ($>$0.8\,M$_\mathrm{Jup}$).
\citet[][]{Burkert-2007} attributed this observation to shorter timescales
of disk depletion for higher-mass stars.

HD\,45652b, a 0.47\,M$_\mathrm{Jup}$ planet with an orbital period of $\sim$44\,days 
orbiting a 0.83\,M$_\odot$ star perfectly fits this scenario.

\begin{acknowledgements}
We thank the Swiss National Research Foundation (FNRS) and the Geneva 
University for their continuous support to our planet search programmes. 
N.C.S. would like to thank the support from Funda\c{c}\~ao para a Ci\^encia 
e a Tecnologia, Portugal, in the form of a grant (references POCI/CTE-AST/56453/2004 
and PPCDT/CTE-AST/56453/2004), and through programme Ci\^encia\,2007
(C2007-CAUP-FCT/136/2006). The support from Coordena\c{c}\~ao de Aperfei\c{c}oamento de Pessoal de N{\'\i}vel Superior (CAPES - Brazil) to R.D.S. in the form of a scholarship is gratefully acknowledged as well.
We acknowledge support from the French National Research Agency (ANR) through project grant NT05-4\_44463.
\end{acknowledgements}

\bibliographystyle{aa}
\bibliography{09402}

\end{document}